# Detecting Communities in Networks by Merging Cliques

Bowen Yan and Steve Gregory
Department of Computer Science
University of Bristol
Bristol BS8 1UB, England
yan@cs.bris.ac.uk, steve@cs.bris.ac.uk

*Abstract*—**Many algorithms have been proposed for detecting disjoint communities (relatively densely connected subgraphs) in networks. One popular technique is to optimize** *modularity*, **a measure of the quality of a partition in terms of the number of intracommunity and intercommunity edges. Greedy approximate algorithms for maximizing modularity can be very fast and effective. We propose a new algorithm that starts by detecting disjoint cliques and then merges these to optimize modularity. We show that this performs better than other similar algorithms in terms of both modularity and execution speed.**

## I. INTRODUCTION

Many systems can be represented as networks: for example, social networks, epidemic networks, neural networks, communication networks, and distribution networks. With the increasing availability of large network datasets, there has been more and more interest in computational techniques that help us to understand the properties of networks. A key topological feature of networks is that vertices are often structured into distinct groups, or *communities*, in terms of edge density. That is, edges between vertices in the same community are dense, but are sparse between different communities [10]. Community detection is important because vertices in the same community often have similar properties. Community detection can allow us to understand attributes of vertices from network topology alone; this has numerous applications in network analysis.

A vast number of community detection algorithms have been developed, especially in the last few years. They vary in the type of network they can handle (unipartite vs. bipartite, weighted vs. unweighted, etc.) and the type of community structure they can detect (disjoint, overlapping, hierarchical, etc.), as well as the techniques used. A very comprehensive recent survey of community detection algorithms appears in [9]. Most algorithms are intended to detect disjoint communities in unweighted, undirected, unipartite networks, and we restrict our attention to this task in the rest of this paper.

Research in community detection has been overshadowed by the lack of a formal definition of *community*. This problem was noted by social scientists decades ago [2] but still has not been solved [9]. Different algorithms tend to adopt different definitions that are consistent with the basic, intuitive concept.

The simplest definition of community is a *clique*, in the graph-theoretic sense: a set of vertices that are all adjacent to each other. In principle, a clique should qualify as a community by any definition, since it is has the greatest possible edge density. There are many existing algorithms for detecting maximal cliques (those that are not subgraphs of other cliques) [5, 28] and maximum cliques (the largest cliques in a network) [14, 22]. However, community detection cannot be reduced to clique detection, for several reasons. First, most communities are not cliques: the requirement that every pair of vertices be connected is too strict in practice. Second, cliques, even if maximal, can be numerous (exponentially related to network size) and highly overlapping. This does not correspond to the intuitive notion of community, especially if communities are expected to be disjoint. Finally, clique detection is computationally expensive: the maximal clique problem and maximum clique problem are both NP-complete.

The first problem has been addressed in the social science literature by relaxing the requirement that every pair of vertices be connected. A community could be an *n*-clique, *n*-club, *n*-clan, *k*-plex, *k*-core, etc., all of which are like cliques but are allowed to have some edges missing. Another approach was adopted in the "clique percolation" algorithm [1, 24]: a k-*clique community* is defined as a set of cliques, each of size *k*, such that each clique has *k*-1 vertices shared with another.

Instead of prescribing properties that a community should satisfy, we can define a function on a subgraph that somehow measures its internal edge density relative to its external edge density, and try to optimize that function. Similarly, we can define a *quality function* that measures the quality of an entire partition (a division of a network into disjoint subgraphs). The most common quality function is *modularity* [20], which has been widely used in recent years to assess the performance of community detection algorithms. It has also been used in some algorithms that work by optimizing modularity, notably [6, 18].

In this paper we propose a simple new community detection algorithm that combines the concepts of cliques and modularity optimization. It consists of two phases: first, we partition the network into a set of disjoint communities which are cliques, favouring larger cliques; second, we merge these communities using a hill-climbing greedy algorithm to maximize the modularity of the partition. As we show, the algorithm achieves better results than other modularity-maximizing algorithms.

In the next section we outline some of the existing related algorithms. Section 3 describes our new algorithm, CliqueMod. In Section 4 we present some results of experiments on synthetic and real networks, to compare our algorithm with related ones. Conclusions appear in Section 5.

## II. RELATED WORK

There are a few existing community detection algorithms that are based on finding cliques: CFinder [1, 24], outlined in Section 1, ComTector [8], described below, and a new algorithm [16] to detect highly-overlapping communities.

The original community detection algorithm based on modularity maximization was Newman's [18] greedy hill-climbing agglomerative algorithm; this was followed by a more efficient version, known as the Fast Modularity algorithm [6]. These algorithms begin with a trivial partition, with very low modularity, in which each vertex is in a separate community. The algorithms then repeatedly merge the pair of communities that results in the greatest increase in modularity, until the desired number of communities is obtained.

The algorithms of [6, 18] give poor results in some cases, because some communities tend to become excessively large. This seems to be because the hill-climbing algorithm has no information at the beginning about which (singleton) communities to merge. Wakita and Tsurumi [27] addressed this problem by modifying the quality function to incorporate balanced community sizes as well as modularity.

An alternative method is to start from a partition in which communities contain more than just one vertex. There are a few algorithms that do this. PBD [26] forms initial communities using random walkers. In PKM [7], each initial community is one of the highest-degree vertices and its neighbours. ComTector [8] first finds all maximal cliques, then converts these to an initial partition by identifying community "kernels" and then assigning each vertex to the most appropriate kernel.

## III. THE CLIQUEMOD ALGORITHM

Our CliqueMod algorithm has two phases:

1. The network is divided into a set of *disjoint* cliques. The size of each clique may be as small as 1, but we prefer larger cliques. These cliques are used as the communities in the initial partition.

2. The number of communities is reduced to the required number by repeatedly merging pairs of communities. The pair chosen at each step is that which maximizes the increase in the partition's modularity.

Because the first phase finds disjoint cliques, it inevitably breaks some large cliques into smaller cliques. For example, if we have a 10-vertex subgraph that is not a clique but contains many overlapping 9-vertex cliques, our first phase will split this into *one* 9-vertex clique and a 1-vertex clique. It is left to the second phase to assemble these into a 10-vertex community.

Our second phase is fundamentally the same as the method used in [6] and some subsequent algorithms. For sparse networks, its time complexity is $O(n^2 \log n)$ in the worst case, but only $O(n \log^2 n)$ if community sizes are balanced, which is more likely with our algorithm than with [6]. Even if the complexity is the same as [6], the execution time will usually be less because fewer merging steps are required.

For the first phase, to find "large" disjoint cliques, the ideal method is to find the largest clique in the network, remove it, find the largest clique in the remaining network, and so on. One way to implement this is to repeatedly use a maximum-clique algorithm, e.g., [22, 14]. An alternative way is to use an algorithm, such as [5, 28], that enumerates *all* maximal cliques and then process its results by repeatedly finding the largest clique and deleting its vertices from the other cliques in the list.

We have found the first option to be faster, because of the exponential number of cliques that need to be examined using the second option. We therefore adopt the maximum-clique algorithm of Konc and Janežič (KJ) [14], with a small optimization: we use the size of the last clique found as an upper bound on the clique size when invoking the maximum-clique algorithm again. (The original algorithm uses graph colouring to obtain an approximate upper bound.) This is the first version of our algorithm, which we name CliqueMod-KJ.

The problem with CliqueMod-KJ is that its time complexity is exponential, because of the use of the maximum-clique algorithm. For large networks, a faster algorithm is essential. We have modified the maximal-clique algorithm of Bron and Kerbosch (BK) [5] so that it finds "approximately maximal" disjoint cliques. Like BK, our algorithm repeatedly expands candidate cliques until they are maximal, but having found one, it now outputs it and immediately deletes its vertices from the algorithm's data structures, instead of exploring alternatives. This way, we have no guarantee that the cliques found are maximal, but they are reasonably large in practice, and there may be more of them, and therefore fewer singleton cliques. This algorithm is much faster, with worst-case time complexity $O(n^2)$ for a sparse network. We use this in the second version of our algorithm, which we name CliqueMod-BK.

Figure 1 shows the clique size distribution in the PGP network [4] after the first phase of CliqueMod-KJ and CliqueMod-BK, respectively.

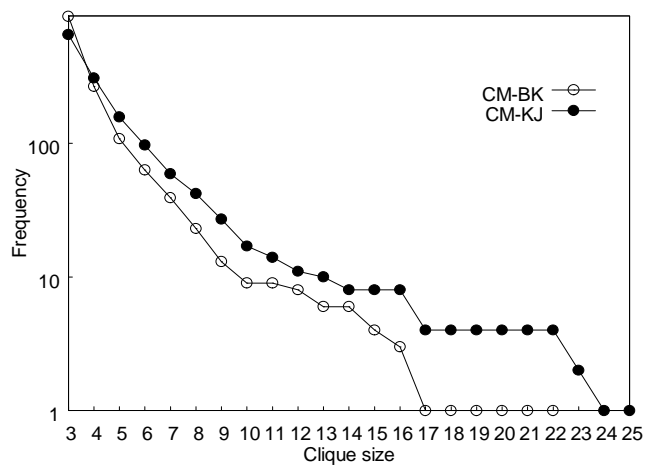

Figure 1. The distribution of clique sizes

## IV. EXPERIMENTS

To evaluate a community detection algorithm, it should be tested on artificial and real networks. For artificial networks, we use the Adjusted Rand index [12] to compare the known communities with the partition found by the algorithm. For real networks, since we do not know the real communities, we use the modularity measure [20] to assess the quality of a partition.

In this section, we evaluate our CliqueMod algorithm (both CM-BK and CM-KJ), implemented in Java, by comparing it with several other disjoint community detection algorithms: Clauset *et al.* (CNM) [6], Wakita and Tsurumi (WT) [27], and Pons and Latapy (PL) [25]. We also compared with the PBD algorithm [26] on real networks, but were unable to do so on synthetic networks because of a run-time error in the PBD code.

### A. Experiments with Synthetic Networks

Lancichinetti *et al.* [15] have proposed a class of artificial networks which they claim reflect the important aspects of real networks, to be used as benchmarks for testing community detection algorithms. The networks have several parameters:

- $n$ is the number of vertices.
- $<k>$ is the average degree. We set the maximum degree to $2.5*<k>$ for all networks we use.
- $\gamma$ is the exponent of the power-law distribution from which the vertex degrees are taken.
- $cs$ is the minimum community size. We set the maximum community size to $2*cs$.
- $\beta$ is the exponent of the power-law distribution from which the community sizes are taken.
- $\mu$ is the mixing parameter: each vertex shares a fraction $\mu$ of its edges with vertices in other communities.

We use these random networks to evaluate the algorithms. All results are averaged over ten random networks generated using the same parameters.

Figure 2 shows the effect of varying the mixing parameter, $\mu$. As $\mu$ increases, CM-BK and CM-KJ are much better than WT and CNM, and slightly better than PL for $\mu$>0.3.

In Figure 3 we vary the community size exponent, $\beta$, over the range 1-2 suggested by Lancichinetti *et al.* [15]. CM-BK and CM-KJ perform better than all other algorithms, and the results seem unaffected by the value of $\beta$. In Figure 4 we vary the degree exponent $\gamma$ over the range 2-3 suggested by Lancichinetti *et al.* [15]. PL is slightly better than CM-BK and CM-KJ, and all three are better than other algorithms. Again, the value of $\gamma$ has little effect on the results.

Figure 5 shows the effect of varying the average degree $<k>$ from 10-30. The maximum degree is set to $2.5*<k>$. PL is slightly better than CM-BK and CM-KJ for higher degrees, but all three are much better than CNM and WT. Figure 6 shows the effect of varying the minimum community size $cs$. The maximum community size is set to $2*cs$. CM-BK and CM-KJ are much better than WT and CNM, and better than PL for larger community sizes.

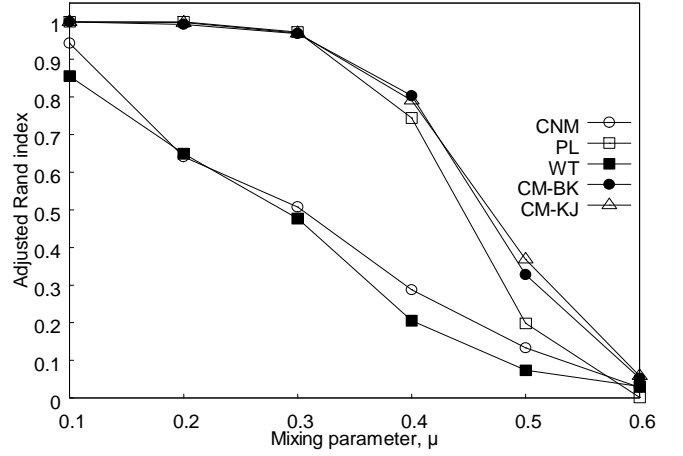

Figure 2. Adjusted Rand index for random networks with $n$=1000, $\beta$=1, $\gamma$=2, $\mu$=0.1~0.6, $<k>$=10, $cs$=50.

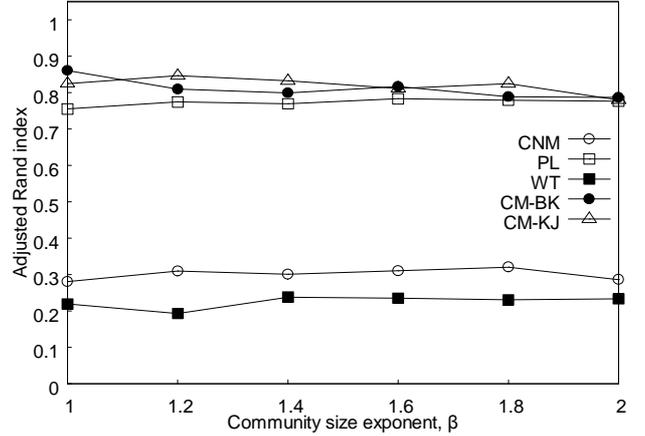

Figure 3. Adjusted Rand index for random networks with $n$=1000, $\beta$=1~2, $\gamma$=2, $\mu$=0.4, $<k>$=10, $cs$=50.

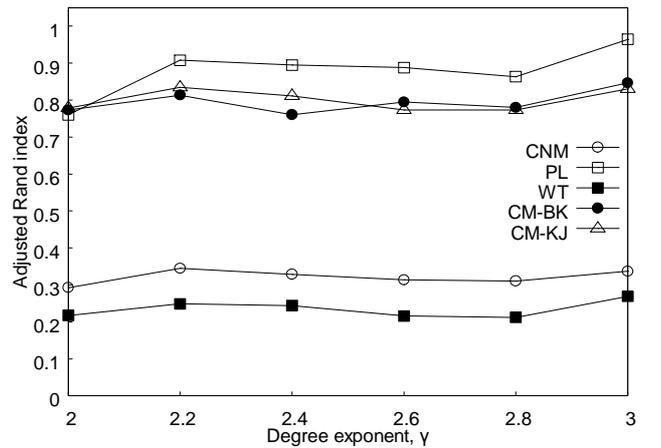

Figure 4. Adjusted Rand index for random networks with $n$=1000, $\beta$=1, $\gamma$=2~3, $\mu$=0.4, $<k>$=10, $cs$=50.

Figure 7 shows how the execution time varies with network size, $n$. All experiments were run on an AMD Opteron 250 at

2.4GHz. As expected, CM-KJ is too slow except for small networks, but CM-BK is faster than CNM and PL.

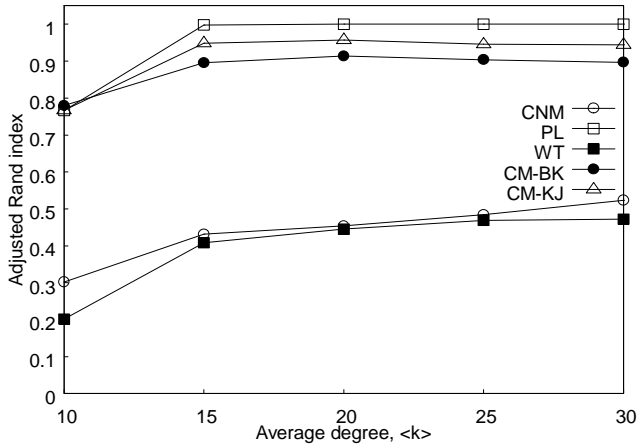

Figure 5. Adjusted Rand index for random networks with $n=1000$, $\beta=1$, $\gamma=2$, $\mu=0.4$, $<k>=10\sim30$, $cs=50$.

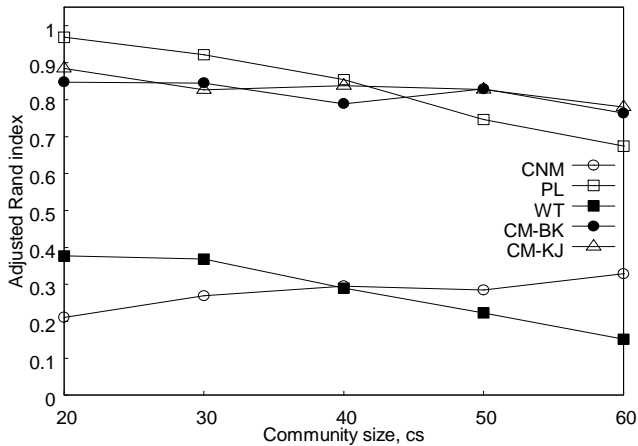

Figure 6. Adjusted Rand index for random networks with $n=1000$, $\beta=1$, $\gamma=2$, $\mu=0.4$, $<k>=10$, $cs=20\sim60$.

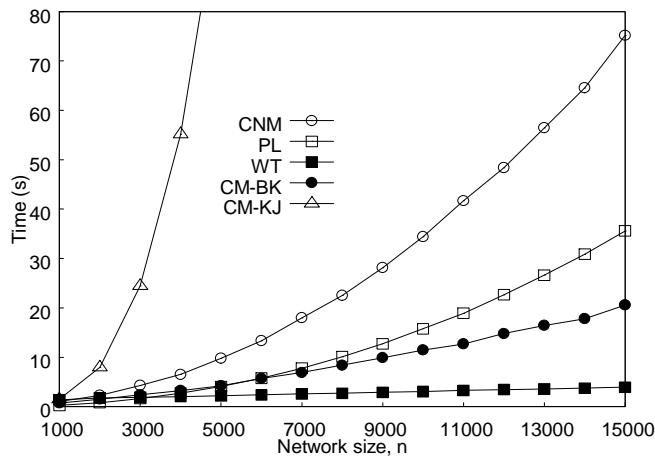

Figure 7. Execution time (seconds) for random networks with $n=1000\sim15000$, $\beta=1$, $\gamma=2$, $\mu=0.4$, $<k>=10$, $cs=50$.

### B. Experiments with Real Networks

We have run the same community detection algorithms on several real networks, listed in Table I. Many of these have become *de facto* benchmarks for community detection algorithms. "scientometrics" is a citation network. "netscience", "cond-mat-2003", "cond-mat-2005", and "erdös1997~2002" are collaboration networks. "c. elegans" is a metabolic network. All others are social networks. Most of the networks used contain only one component, but "drugnet", "erdös1997~1999", and "cond-mat-2005" contain several components, for consistency with the experiments reported in [7, 8].

Table I shows the maximum modularity obtained by each algorithm for each of these networks, and the total execution time on an AMD Opteron 250 at 2.4GHz. As well as CliqueMod, CNM, WT, PL, and PBD [26], the table shows the results for the ComTector (CT) [8], and PKM [7] algorithms, as reported in the respective papers. Regrettably, we have been unable to obtain the code for CT and PKM, so our comparison of modularity is incomplete, and the execution times are not directly comparable with the others.

Our results show that CliqueMod-KJ generally gives the highest modularity, closely followed by CliqueMod-BK, while both of them give better solutions than the other algorithms. For the larger networks, CM-BK is the fastest algorithm tested, with the exception of WT and (possibly) PBD.

## V. CONCLUSIONS

We have presented a simple algorithm to detect disjoint communities in networks by merging cliques. Although large cliques are desirable, they need not be maximal: our approximate algorithm, CliqueMod-BK, gives results that are almost as good as CliqueMod-KJ, which uses maximal cliques, and almost always better than those of the other algorithms. Our algorithm is among the fastest, though not as fast as the WT algorithm, which gives consistently lower modularity.

It is perhaps not surprising that our algorithms' results are better than those of the CNM and WT algorithms, since those begin the merging process with singleton communities, as we noted in Section 2. More interesting is the comparison with the algorithms of [7, 8, 26], which also start from larger initial communities. We believe that our use of cliques works better, in modularity terms, because cliques are as the most modular initial communities possible. Although ComTector [8] also finds cliques, the initial communities are created from them by a process that seems to *reduce* modularity.

One area of future work is to improve our algorithm's speed even further. Our current implementation is just an unoptimized prototype. It may be worth applying the same techniques that the WT algorithm adds to the CNM algorithm to balance community sizes and improve speed. Other future work includes extending our algorithm to allow overlapping communities. This is not trivial, because of the massive overlap that already exists between cliques and the difficulty in optimizing the modularity of overlapping communities [21].

Our CliqueMod implementation is available from http://www.cs.bris.ac.uk/~steve/networks/cliquemod/ .

TABLE I.    REAL NETWORKS: MAXIMUM MODULARITY AND EXECUTION TIME

| Name | Ref | Vertices | Edges | Maximum modularity | | | | | | | | Execution time (s) | | | | | | |
|---|---|---|---|---|---|---|---|---|---|---|---|---|---|---|---|---|---|---|
| | | | | CliqueMod | | CNM | PL | WT | PBD | CT* | PKM* | CliqueMod | | CNM | PL | WT | PBD | CT* |
| | | | | BK | KJ | | | | | | | BK | KJ | | | | | |
| zachary | 30 | 34 | 78 | 0.417 | 0.417 | 0.387 | 0.335 | 0.419 | 0.394 | | 0.412 | 0.03 | 0.04 | 0.05 | 0.01 | 0.55 | 0.004 | |
| drugnet | 7 | 212 | 284 | 0.747 | 0.745 | 0.745 | 0.702 | 0.724 | 0.735 | | 0.751 | 0.05 | 0.05 | 0.20 | 0.03 | 0.93 | 0.01 | |
| netscience | 19 | 379 | 914 | 0.835 | 0.847 | 0.837 | 0.828 | 0.819 | 0.837 | | | 0.06 | 0.20 | 0.37 | 0.02 | 0.75 | 0.01 | |
| c. elegans | 13 | 453 | 2025 | 0.430 | 0.418 | 0.408 | 0.349 | 0.421 | 0.416 | | | 0.04 | 0.26 | 0.43 | 0.07 | 0.94 | 0.04 | |
| email | 11 | 1133 | 5451 | 0.550 | 0.554 | 0.512 | 0.531 | 0.478 | 0.536 | | | 0.38 | 1.99 | 1.21 | 0.26 | 1.43 | 0.06 | |
| scientometrics | 23 | 2678 | 10368 | 0.596 | 0.608 | 0.548 | 0.533 | 0.517 | 0.560 | | | 1.11 | 18.6 | 2.92 | 1.01 | 1.79 | 0.22 | |
| blogs | 29 | 3982 | 6803 | 0.857 | 0.857 | 0.850 | 0.788 | 0.832 | 0.839 | | | 0.84 | 36.9 | 3.80 | 0.65 | 7.47 | 0.37 | |
| erdös1997 | 3 | 5482 | 8972 | 0.727 | 0.733 | 0.523 | 0.679 | 0.712 | 0.709 | 0.69 | | 2.75 | 33.8 | 5.80 | 2.24 | 2.45 | 0.72 | 23 |
| erdös1998 | 3 | 5816 | 9505 | 0.727 | 0.734 | 0.706 | 0.676 | 0.711 | 0.714 | 0.69 | | 2.92 | 38.3 | 6.19 | 2.50 | 2.66 | 0.82 | 26 |
| erdös1999 | 3 | 6094 | 9939 | 0.732 | 0.733 | 0.699 | 0.678 | 0.709 | 0.714 | 0.69 | | 3.29 | 42.1 | 6.53 | 2.76 | 2.61 | 0.86 | 27 |
| erdös2002 | 23 | 6927 | 11850 | 0.702 | 0.697 | 0.670 | 0.627 | 0.676 | 0.682 | | | 4.32 | 58.4 | 7.67 | 5.32 | 3.72 | 1.36 | |
| PGP | 4 | 10680 | 24316 | 0.877 | 0.877 | 0.855 | 0.789 | 0.851 | 0.861 | | | 5.06 | 857 | 11.1 | 3.72 | 2.88 | 1.78 | |
| cond-mat-2003 | 17 | 27519 | 116181 | 0.734 | 0.748 | 0.657 | 0.629 | 0.709 | 0.725 | | | 36.6 | 26476 | 127 | 59.2 | 5.22 | 10.4 | |
| cond-mat-2005 | 17 | 39577 | 175693 | 0.707 | | 0.655 | 0.593 | 0.577 | 0.689 | 0.65 | | 106 | | 278 | 138 | 6.75 | 29.0 | 7920 |

*Results reported in original papers, not tested by us.